# How Charles Babbage Invented the Computer

Raúl Rojas
Department of Mathematics and Statistics
University of Nevada Reno

Abstract: This paper provides an overview of the successive stages in the development of Charles Babbage's Analytical Engine, based on the blueprints held in the Babbage Papers Archive, accessible online through the Science Museum in London. The first person to decipher these schematics was Allan Bromley, whose contributions in the 1980s and 1990s significantly advanced our understanding of Babbage's pioneering work. The Science Museum's digitization of the Babbage Papers enables a chronological exploration of the evolution of Babbage's machines. The focus is on the Analytical Engine, shedding light on its lesser known but crucial transitional phases.

## 1 Introduction

The English mathematician and scholar Charles Babbage (1791-1871) designed the first automatic computing machines almost 200 years ago [Collier 1970]. He drew inspiration from the work of Gaspar de Prony in France, who many years earlier had tackled the task of producing mathematical tables through the implementation of a numerical assembly line [Campbell-Kelly et al. 2003, Babbage 1832]. The methodology involved a group of mathematicians who chose a polynomial to serve as an approximation to a target function. Then the "method of differences" was applied. This entailed having a tabular representation for the function to be approximated and its differences. Adding and subtracting the differences produced a table of successive function values. All that was needed were workers with basic arithmetic skills (in the 19th and early 20th century such individuals were called “computers” [Swade 2003]).

| $n$ | $n^2$ | 1st difference | 2nd difference |
|---|---|---|---|
| 1 | 1 | | |
| 2 | 4 | 3 | |
| 3 | 9 | 5 | 2 |
| 4 | 16 | 7 | 2 |
| 5 | 25 | 9 | 2 |

Table 1: The computation of $n^2$ using a table of differences

Consider Table 1, which outlines the procedure for computing the squares of integers, starting from $n = 1,2,3$, and so forth. The initial three rows of the table are computed manually. The difference between $(n+1)^2$ and $n^2$, for consecutive values of $n$, constitutes the initial difference. The subsequent differences of the first differences are recorded in the column labeled second differences, and so on. When the difference becomes constant (in this instance, settling at 2), we can generate new rows for the table by summing the differences from right to left, thus forming the additional rows. This method is applicable to the computation of any polynomial. Given that many functions can be approximated by polynomials (via their Taylor series), this straightforward technique furnishes a means to construct tables for various mathematical functions. Periodically, it is imperative to initiate the process anew from a different approximation point in order to maintain the error of the Taylor series within acceptable bounds.

In 1819, Babbage began to seriously contemplate the automation of this repetitive task [Hyman 1982], although he had been mulling over machines for mathematical tables since 1813 [Babbage 1864]. He realized that the process of filling the table by the method of differences could be completely automated if each column in the table was a decimal accumulator, capable of receiving a decimal number, adding (or subtracting) it to its previous content, and then passing the result to the next accumulator. Such a linear array of accumulators would constitute the Difference Engine (DE) that Babbage began to design and for which he received government funding. Fig. 1 shows an engraving of the small model of the DE that Babbage built between 1820 and 1822 (as a proof of concept [Babbage 1864]). The vertical accumulators are visible. Each accumulator consists of a column of six gears, capable of storing decimal numbers of up to six digits.

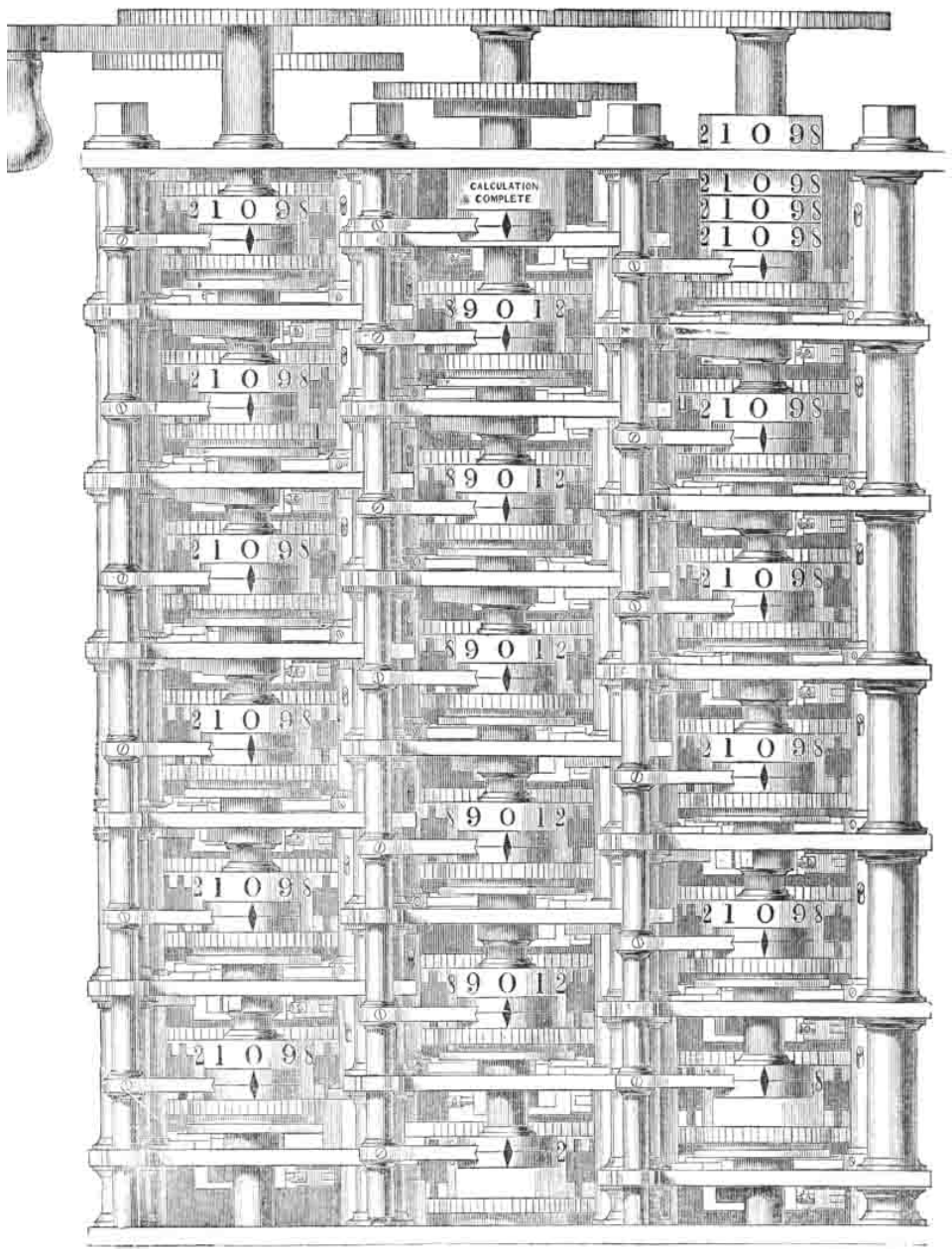

Fig. 1: Frontispiece of *Passages of the Life of a Philosopher* [Babbage 1864]

## 2 The Analytical Engine

Babbage never finished building his first Difference Engine (there was a second, later on), mainly because he never really stopped improving the design and never settled on a definitive set of blueprints for the ultimate machine. While working on the DE, he actually had a better idea: if the results of the machine could be fed from the last column, where the mathematical function is being produced (column 1 in Table 1), to the last difference considered (last column on the right in Table 1), then functions more complex than polynomials could be computed. Babbage called this idea "the machine that eats its own tail" [Babbage 1864]. From then on, Babbage concentrated his energy on designing the new calculating machine, which he christened the "Analytical Engine" (AE). The name reflected the fact that the machine should be able to compute all the interesting functions of analysis. As Babbage put it, “the whole of arithmetic now appeared within the grasp of mechanism” [Babbage 1864].

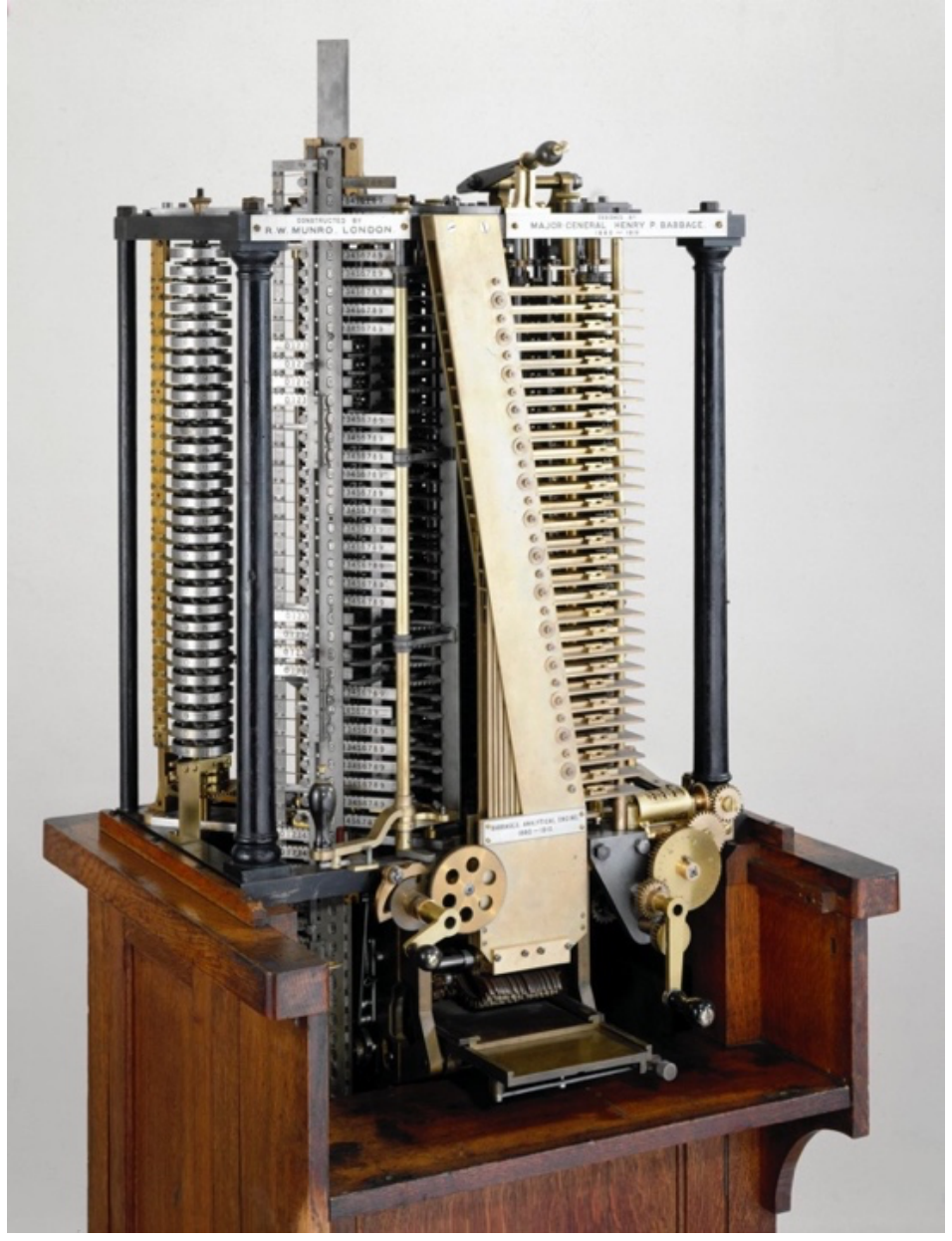

Fig. 2: Model of the mill of the Analytical Engine built by Charles Babbage’s son (Science Museum, Object 1896-58, CC BY-NC-SA 4.0). Nery Prevost Babbage also published the first collection of papers related to the Analytical Engine [Babbage 1889].

Allan Bromley's seminal studies [1987] provided a chronological framework for Charles Babbage's work on the Analytical Engine. The first thing to emphasize is that Babbage never really stopped developing the Analytical Engine, and that no serious effort was made to actually build the complete machine (until a small demonstration model was planned). In 1910, Henry Prevost Babbage, Charles Babbage’s son, built a scaled down demonstration model (Fig. 2). The existing records of the Analytical Engine can thus be viewed as an unfulfilled vision of how a computer might be built, a vision embodied in various "plans" (drawings) of the machine. Among these, Plan 25, is the most famous of them all, and is often referred to when discussing the Analytical Engine. However, it was not the final proposal, and there are significant differences between the early designs and the later ones, especially for the prototype that was partially built after Babbage's passing. Much like later computing pioneers, Babbage gradually streamlined his designs, which progressively came closer to the architecture of modern computers. There is an economy of means that characterizes the evolution of the various plans for the Analytical Engine. In 1864, Babbage wrote: "The great principles on which the Analytical Engine rests have been examined, admitted, recorded, and demonstrated. The mechanism itself has now been reduced to unexpected simplicity" [Babbage 1864]. That simplicity is clear in Fig. 2, a small model of part of the processor of the AE constructed by Babbage's son.

This is the story that we want to review in this paper, namely the evolution of design.[1] Let us start by examining Bromley's proposed timeline for Babbage's work on computing machines. Fig. 3 shows the main periods of development of the Analytical Engine along some pivotal milestones. The chronicle begins in 1833 when Babbage decided to build a more general-purpose computing machine. Plans for the AE swiftly took shape, starting in 1834. Work continued with several interruptions until 1846, when Babbage decided to redesign the Difference Engine (which then became the DE number 2 [Swade 2001]). The Babbage Papers at the Science Museum also contain a collection of drawings catalogued as the "second phase of work" on the Analytical Engine, covering the period from 1857 to Babbage's death in 1871. Bromley identified five main periods of development up until 1846 (shown as brackets on the right in Fig. 3). Adding the work done between 1857 and 1871, we can extend Bromley's chronology to six major periods.

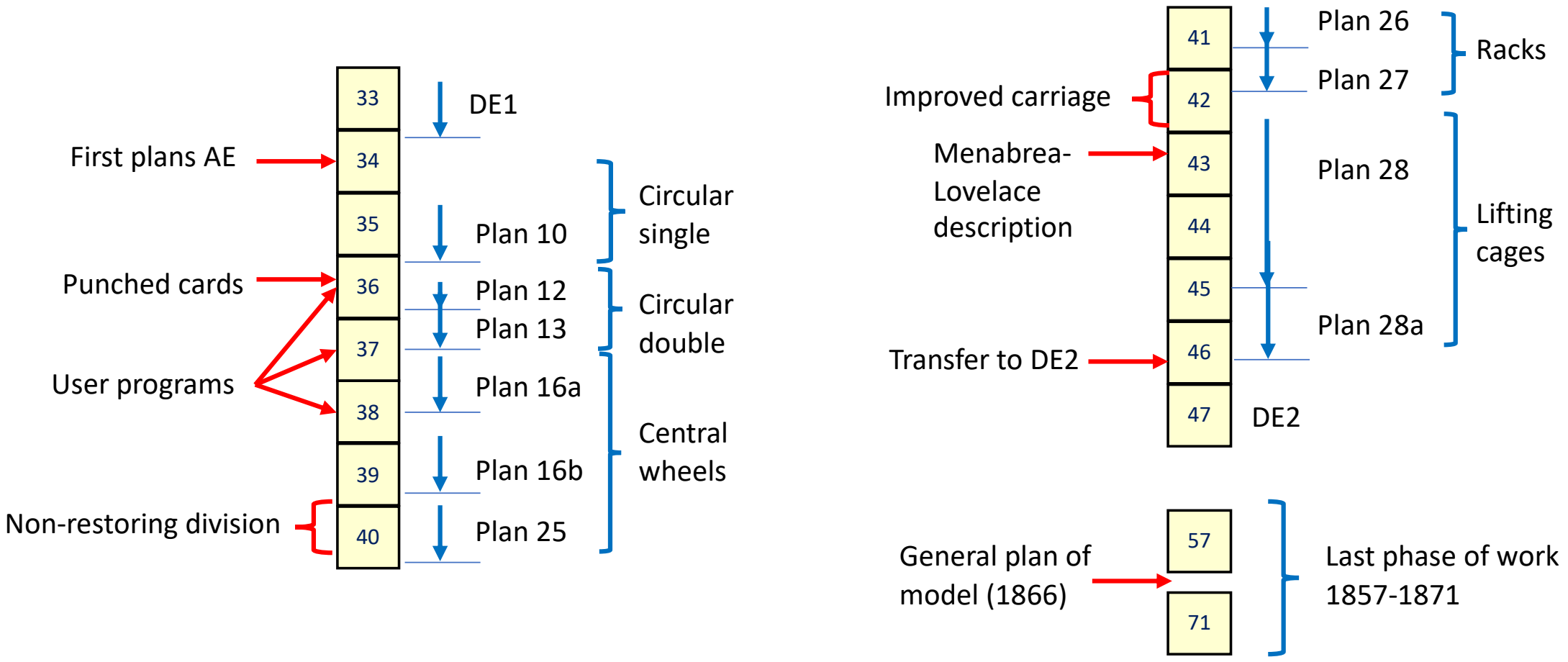

Fig 3: Timeline of Babbage's work on the Analytical Engine. Based on [Bromley 1987] and Babbage Papers Archive

The original idea for the Analytical Engine evolved continuously from 1833 to 1846. After a few years of refinement, the design coalesced around features that we now recognize as characteristic of modern computers [Babbage 1837]. The Difference

[1] The definitive account of the evolution of the Analytical Engine was written by Bromley [1987]. Here we want to touch on some points not covered by him, specially regarding the architecture of the 1866 small model of the AE. We also provide some enhanced illustrations of the various plans.

Engine consists of accumulators linked in a chain, blurring the distinction between processing and memory. An accumulator can hold a decimal number, but it can also add or subtract another number arriving from the adjacent accumulator. In contrast, the Analytical Engine featured a "store" for variables, i.e., decimal numbers, and a separate "mill" for processing the four basic arithmetic operations. For example, if we want to add two numbers, they are sent from the memory to the mill, and the result is transcribed back from the mill to one of the "variables" (addresses) in the memory. Each variable in the store should be able to hold decimal numbers with many digits (Babbage considered 50, 40, and also 25 digits per variable, in different drafts). Each decimal digit was stored in a gear, which could encode the digits 0 to 9 according to its angle of rotation relative to the zero reference, as traditional mechanical calculators could do. With $n$ gears aligned along the same vertical axis, a column of $n$ digits could be built, i.e., a decimal number with the same number of digits could be stored there (see Fig. 1).

Crucial for the history of the Analytical Engine is the fact that Babbage moved away from a "self-contained" design (as in the case of the Difference Engine) in which the machine would only read numbers and execute a program "hardwired" on mechanical barrels, akin to the mechanics of drum music boxes. But instead of producing notes, the pins on the rotating barrels would set in motion levers that orchestrated all the mechanical operations in the machine [Bromley 1998]. The “programmer” would have to reposition the pins on the master barrel according to the function to be computed. Around 1836, however, Babbage decided to use Jacquard cards instead, which at that time were used to specify the textures woven by mechanical looms. Now, the high-level instructions for the computation would be held in such punched cards. The stream of operation cards constituted the "program" to be executed by the mill. In addition, he envisioned a stream of cards for the addresses of the variables involved in the computations, and yet another stream for numbers that could be fed into the machine's memory.

Thus, at a very early stage, after only three years of work on the Analytical Engine, an audacious proposal emerged for a device that

- would implement the separation of processor and memory typical of today’s so-called von Neumann computers,
- whose processor and memory would be coordinated by an external program housed on punched cards,
- and where the machine would be "microprogrammed", that is, its elementary movements would be deftly orchestrated by drums that would sequence the microoperations necessary for each arithmetic operation.

The main problem now is the strategic arrangement of the various components of the machine, and here we find the five main periods of development identified by Bromley until 1846:

- The initial drafts of the AE describe a machine built around a circular master gear (a sort of "bus") that would transfer numbers between the different parts of the machine.
- In the second phase (Plan 12 and Plan 13), a machine with a double mill is proposed, that is, two processors fashioned as mirror images that can work independently, or in tandem.
- In the third phase, the mill retains its central wheel, while the store gets its own linear bus (called a "rack" by Babbage).
- In the fourth phase, there is an important simplification when the AE is designed only with linear racks, rendering the circular bus obsolete. This made it possible to further compact the positioning of the various elements of the machine.
- In the fifth phase, up to 1846, Babbage further simplified by eliminating many auxiliary gears, using instead columns of gears that could be raised or lowered to engage with other components of the AE.

Each new plan for the AE is simpler than the previous one, culminating in Plan 28 and Plan 28a by 1846, with another partial design for a demonstration machine in 1866 (during the sixth and final period of development). In the following sections we will discuss each of these phases.

## 3 The circular machines

In the initial phase of Analytical Engine development, a circular framework took center stage, a nod to the original concept of arranging the accumulators in a closed chain. However, each accumulator was expensive, in the sense that it not only had to store a decimal number of many digits, but it also had to be able to add new incoming data by propagating the decimal carries. The inherent complexity of having many accumulators would make the machine expensive and slow. Babbage's solution was twofold: on the one hand, he decided to have specialized mechanical columns that would be used only for storing numbers but not for computing (and so the "store" was conceived). On the other hand, the execution of the arithmetic operations was consolidated in specialized components, and so the "mill" was born. A large circular gear was needed in order to move information from one component to another. Actually, the circular gear consists of a column of parallel gears, arranged like a cylinder, since for each digit in each decimal column a way to transport each digit is needed. Babbage's drawings show the AE from above, in a vertical view, and this must be taken into account when interpreting them. One gear, seen from above, can represent a column of 40 or 50 gears (Babbage called such columns "axes").

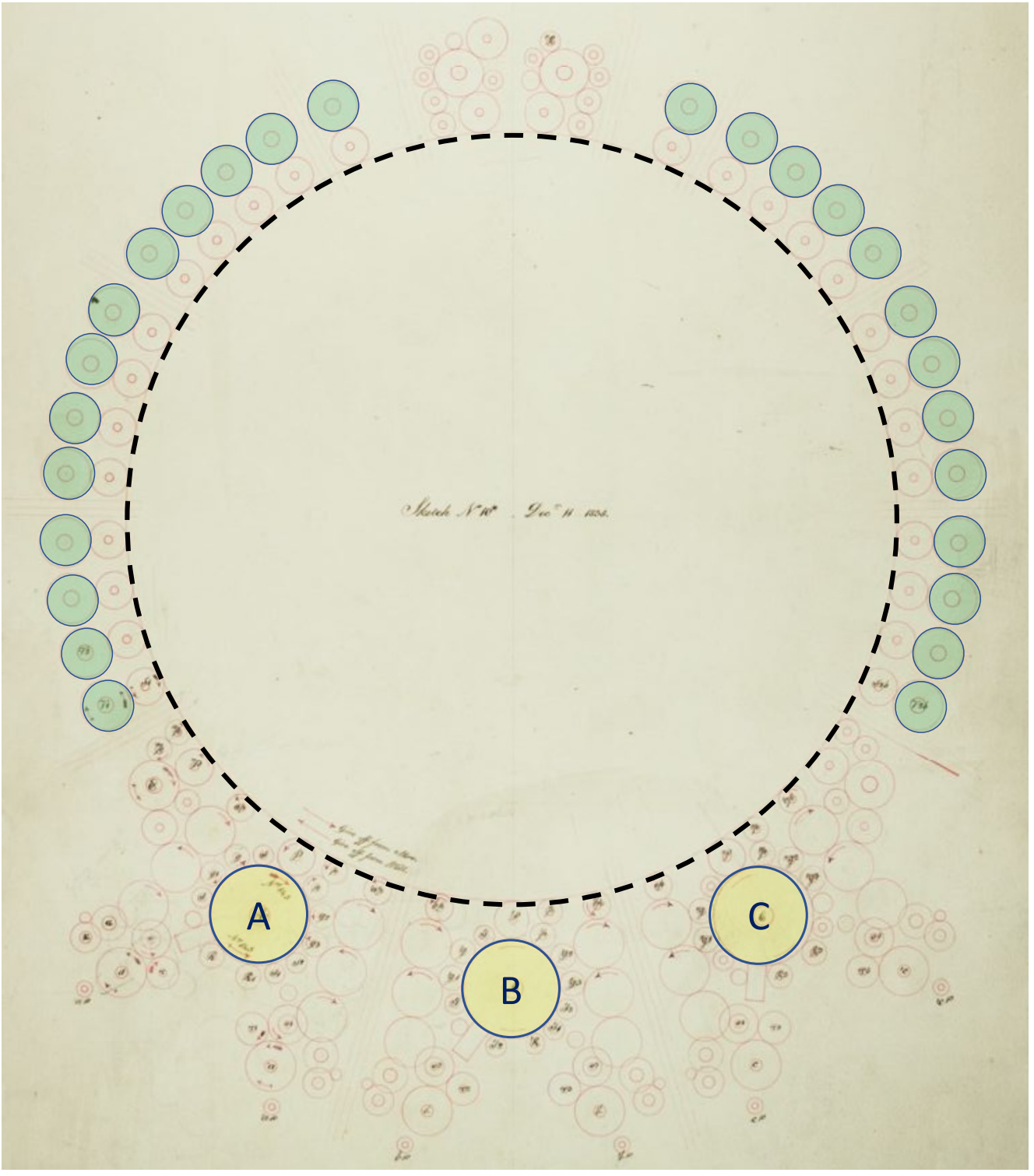

Fig. 4: Plan 10: Babbage Papers, BAB/A/026 of December 11, 1835

Fig. 4 shows Plan 10 of 1835 (one year into the project). Here, the circular configuration of all elements is prominently displayed. The upper part shows the store variables (highlighted in green) linked to the circular gear by intermediate gears. The lower part shows the arrangement of the elements of the mill. Three accumulators are visible, marked with the letters A, B, C. The gears encircling the accumulators manage the decimal carries. It is also possible to "shift" the accumulators up or down, which can then be multiplied or divided by ten. It is important to note the striking simplicity of the design and the intended reuse of components. However, many details of the later machines are still to be fleshed out.

Plan 13 of 1837 (Fig. 5) retains the circular structure, but now the mill has been designed with two symmetrical parts that mirror each other. The barrels for microprogramming are conspicuously present (for example, in the upper left corner), while the components on the far left are the readers for the punched cards. Babbage wrote the first programs for the AE in 1836/37, by which time he felt confident enough in the design to begin writing code for it [Rojas 2023]. Bromley dates the introduction of punched cards for the AE to June 1836 [Bromley 1987]. The large circular gear serves still as a linchpin, connecting all components. Plan 13 is much more detailed than Plan 10, and now the difficulty of connecting and arranging all the components becomes apparent. Apparently, Babbage conceived the dual-mill AE as a machine capable of working on two different computations at the same time. However, this would have made the control part of the machine too complicated. At the farthest right end of Plan 13 we can see the printing apparatus.

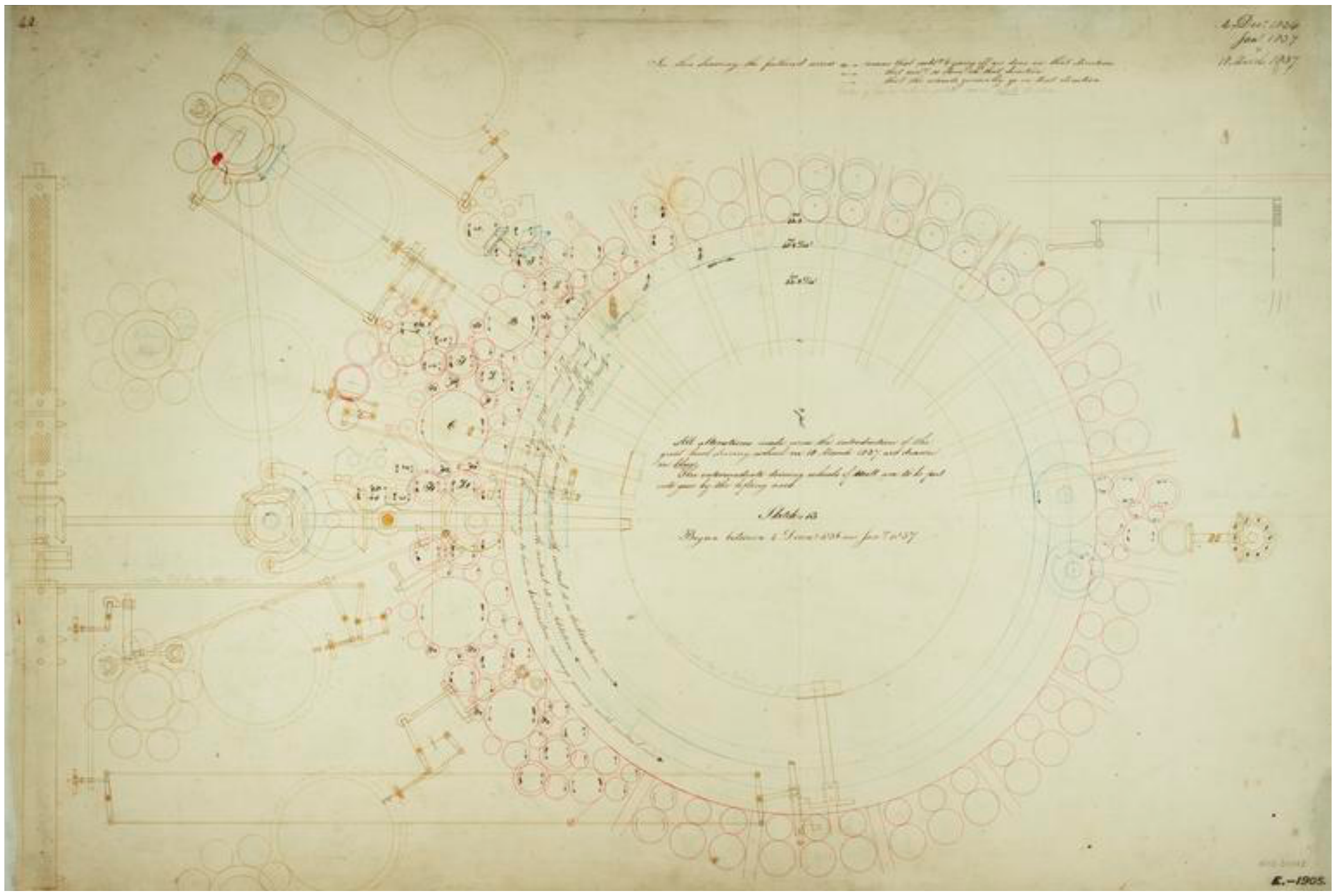

Fig. 5: Plan 13 of March 1837 (Babbage Papers BAB/A/042)

Bromley [1987] published an enhanced rendition of Plan 13, where the different parts of the AE are more readily discernible. Mill 1 and Mill 2, in his drawing (Fig. 6), each feature three figure axes (accumulators), two of them with anticipating carry. There are six barrels for microprogramming the mill.

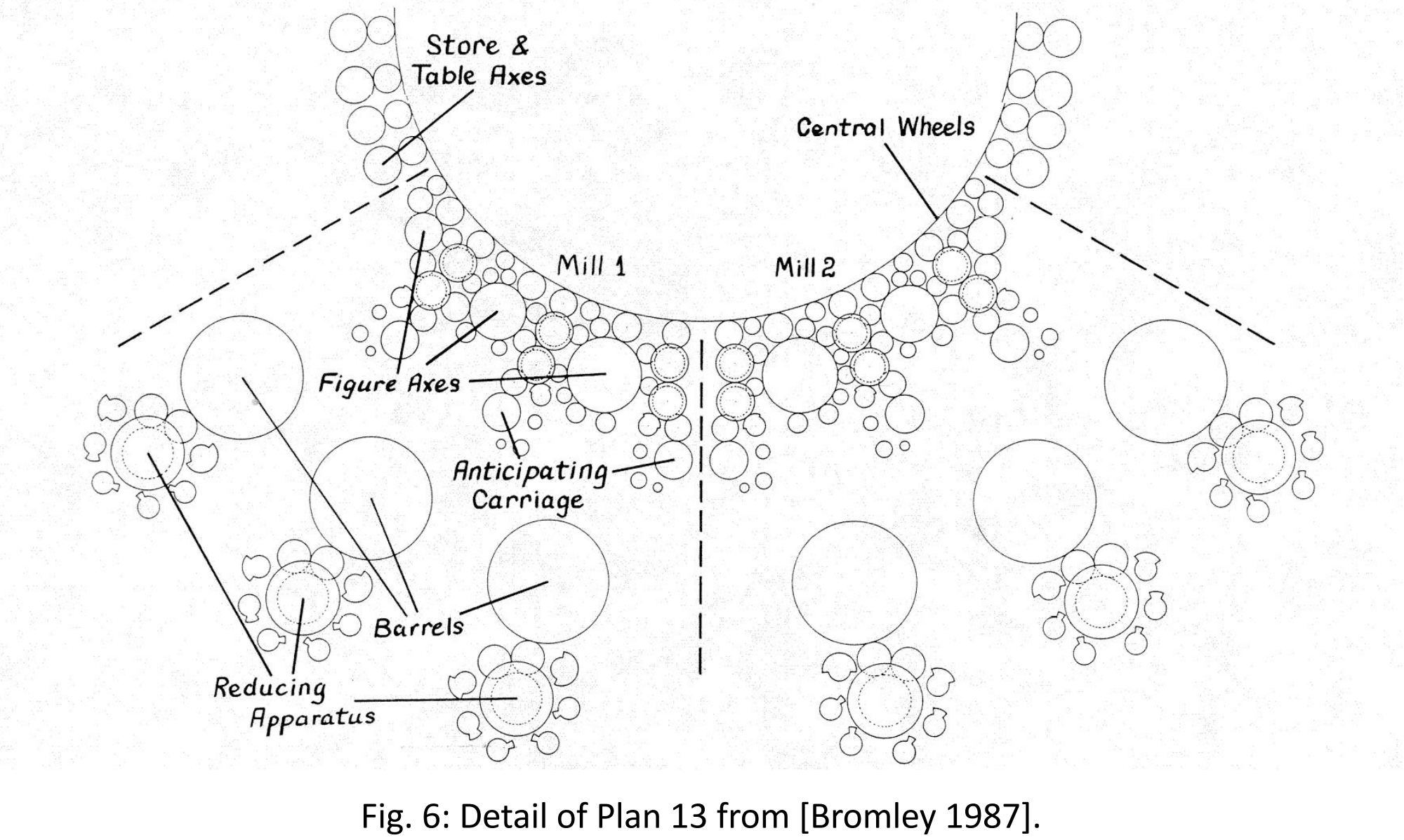

Fig. 6: Detail of Plan 13 from [Bromley 1987].

Fig. 7 shows a diagram of the type of punched cards and reading prism that Babbage intended to use for the AE. The cards were threaded together and contained the high-level commands executed by the microprogramming drums.

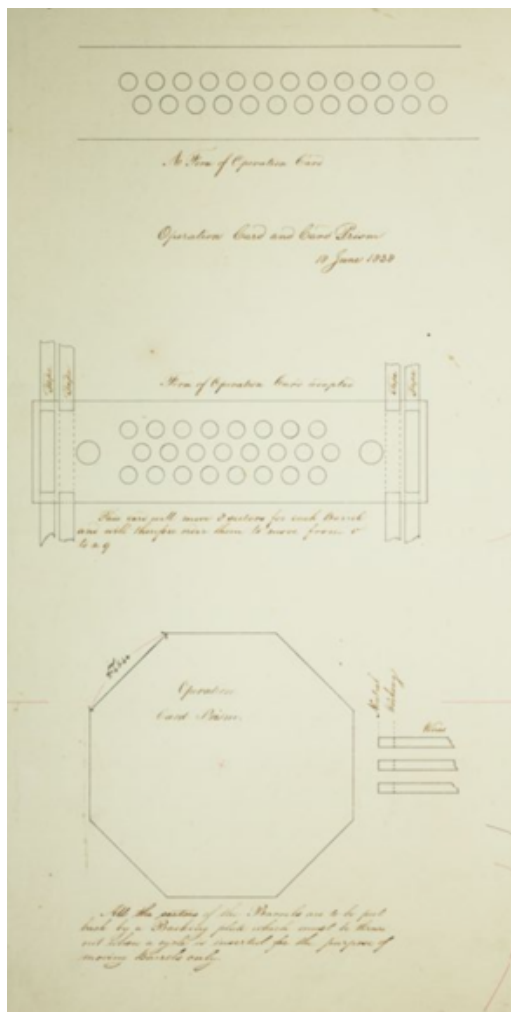

Fig. 7: The punched operation cards and the prism for reading them (BAB/A/064 of Sept. 17, 1839)

## 4 Mill with central wheels and store with racks: Plan 21 and Plan 25

From 1837 on, Babbage modified his design so that a central wheel would act as the bus for the mill and a linear rack would act as the bus for the store. Plan 21, dated November 1839, shows clearly the new distribution of computing elements. The accumulators A and ‘A (highlighted in green) are each coupled with a carry unit (F and ‘F, highlighted in blue). Additionally, a third carry unit, ‘’F, works with the ingress axis I (another accumulator, highlighted in green). Information flows from the store to the mill through the ingress axis I. The results flow back from the mill to the store through accumulator ‘’A (also in green, to the right).

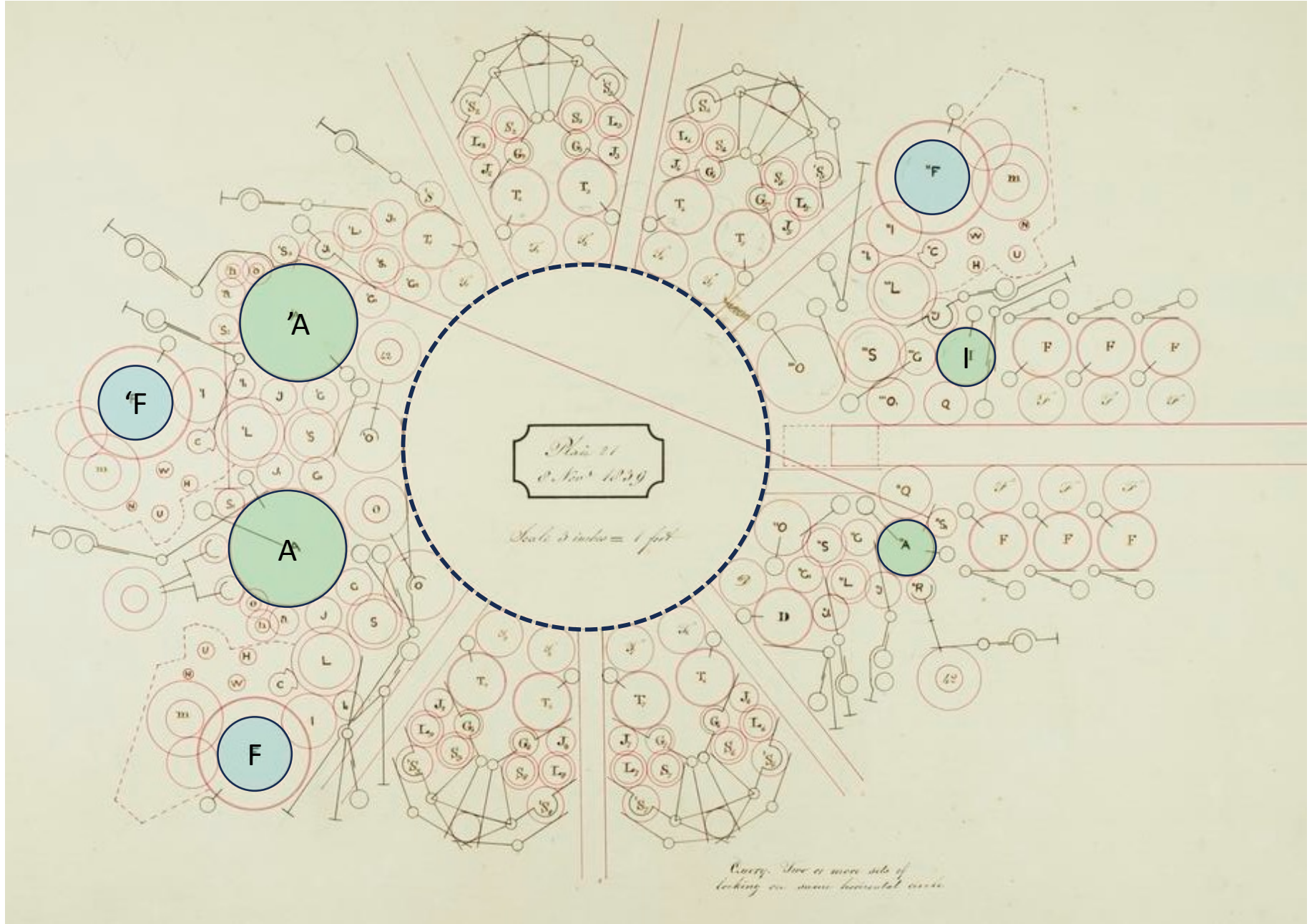

Fig. 8: Plan 21, November 1939 (Babbage Papers BAB/A/076)

There are 9 gears marked with T. They store the multiples of the multiplicand, for each digit from 1 to 9. The multiplier, in turn, is kept in accumulator ''A. Suppose that we want to multiply 32 by 16. First, the mill computes the multiples $1 \times 32, 2 \times 32, \dots, 9 \times 32$ and stores each one in the axis T1 to T9. Then, when multiplying by 16, the content of axis T6 is added to accumulator A. Then, axis T1 is shifted up one position (which multiplies the multiplicand by 10) and the result is added to accumulator A. The final result is 10×T1+T6, that's is, 16×32. This is the algorithm for "multiplication with a table" that Babbage decided to implement. The axes T1 to T9 survived until the last designs for the AE, with the difference that in later designs the T-axes were not shifted up. Instead, it was the accumulators A and 'A which were shifted down after each partial addition, so that the final result would be the same, but all the machinery for shifting the T-axes could be spared.

By August 1840, Babbage had arrived at his tentative final design, which he then ordered engraved. Now all elements of the AE are visible on a single plate (Fig. 9).

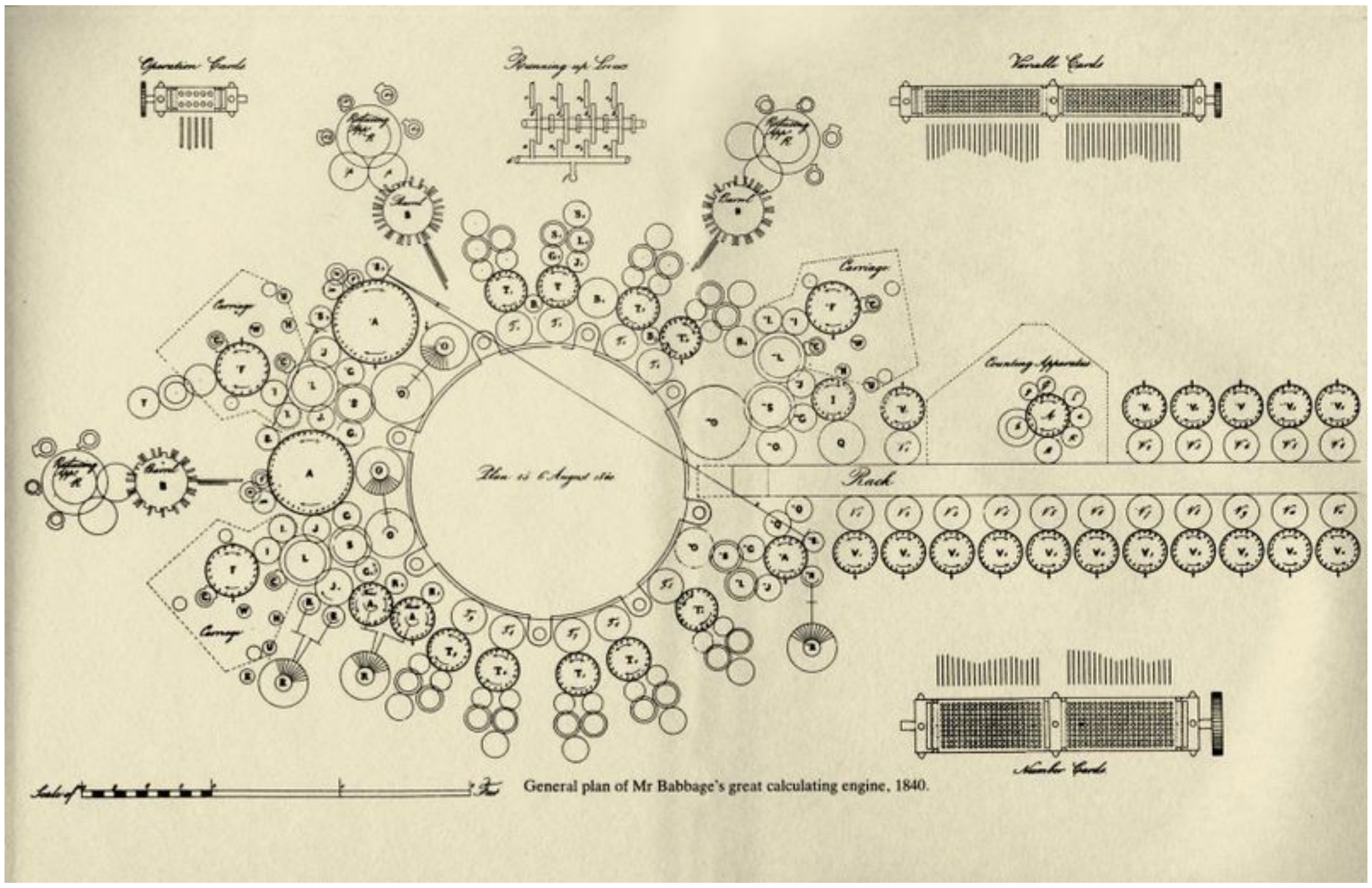

Fig. 9: Plan 25, August 1840 (Babbage Papers BAB/A/089)

In Plan 25, the radius of the central mill wheel is a little more than one foot, while the entire mill's diameter spans approximately six feet. Very prominent are the three sets of punched cards: on the left we have the operation cards and their reader. On the top right are the variable cards, which were read in two streams that could move independently. For the number cards we also have two streams, and they could also move independently of each other, to produce different pairings of cards.

It is easier to understand Plan 25 if we look at the same drawing, with the different parts colored according to their function.

Fig. 10 shows the highlighted components of the AE, according to Plan 25. On the left we find the variables constituting the store. Each blue circle represents a column of storage gears with the same axis. The yellow circles represent gears that connect the variables to the linear bus (the rack). The yellow gears can engage or disengage the bus as needed. Some other yellow circles also represent gears that connect other components to the circular bus in the mill.

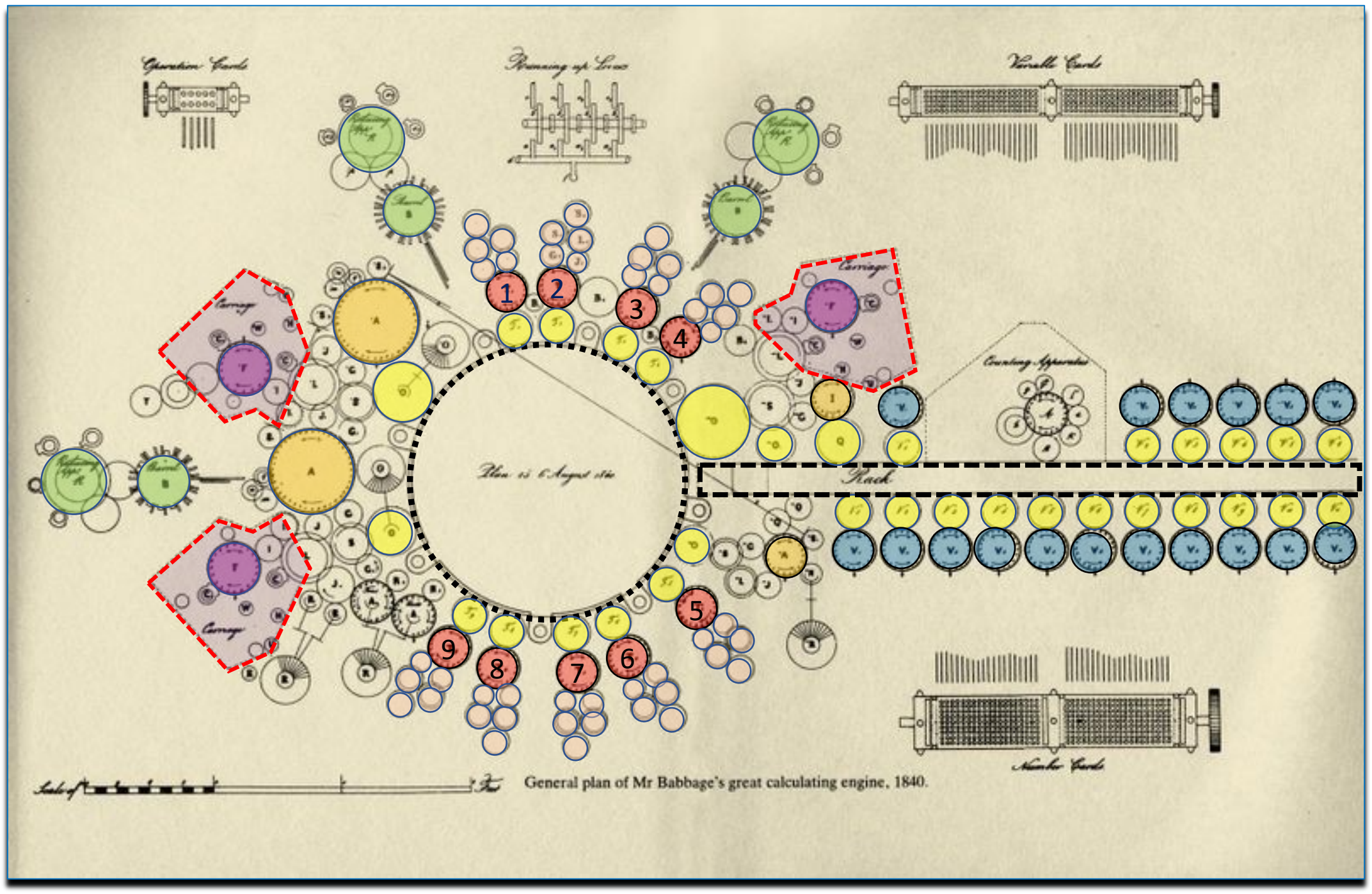

Fig. 10: Plan 25, with highlighted components

The red circles represent the T-axes that hold the multiples of the multiplicand, one for each decimal digit, from 1 to 9. These multiples connect to the circular bus by yellow gears. The orange circles represent accumulators for storing numbers. The main accumulators are A and ‘A (on the left). They can work separately or in tandem, doubling the digit capacity. There is one accumulator (labeled I, for ingress) that holds variables moving from the store to the mill. Another accumulator (‘’A), near the store, holds results in transit from the mill to the store.

The purple boxes and gear are the carry look-ahead units. There is one for each accumulator A and ‘A, and one for the ingress axis I. The green circles represent the barrels used for microprogramming the AE.

## 5 The linearized Analytical Engine: Plans 26 to 28

Babbage persisted in refining the AE beyond Plan 25, resulting in an architecture that transitioned into a linearized form, effectively obviating the central wheel. The arrangement of the different components is now easier to grasp. Fig. 11 shows the main details of Plan 26. Highlighted in blue we can see the three accumulators needed for multiplication (A and ‘A), as well as the accumulator needed to hold the multiplier during the multiplication algorithm. The accumulators A and ‘A have associated carry units (F and F’). The three accumulators interface seamlessly with the horizontal rack, which connects on top with the table of multiples of the multiplicand (represented by a single axis T) and the store variables V. The meaning of the different labels is explained on the right (unreadable at this resolution level).

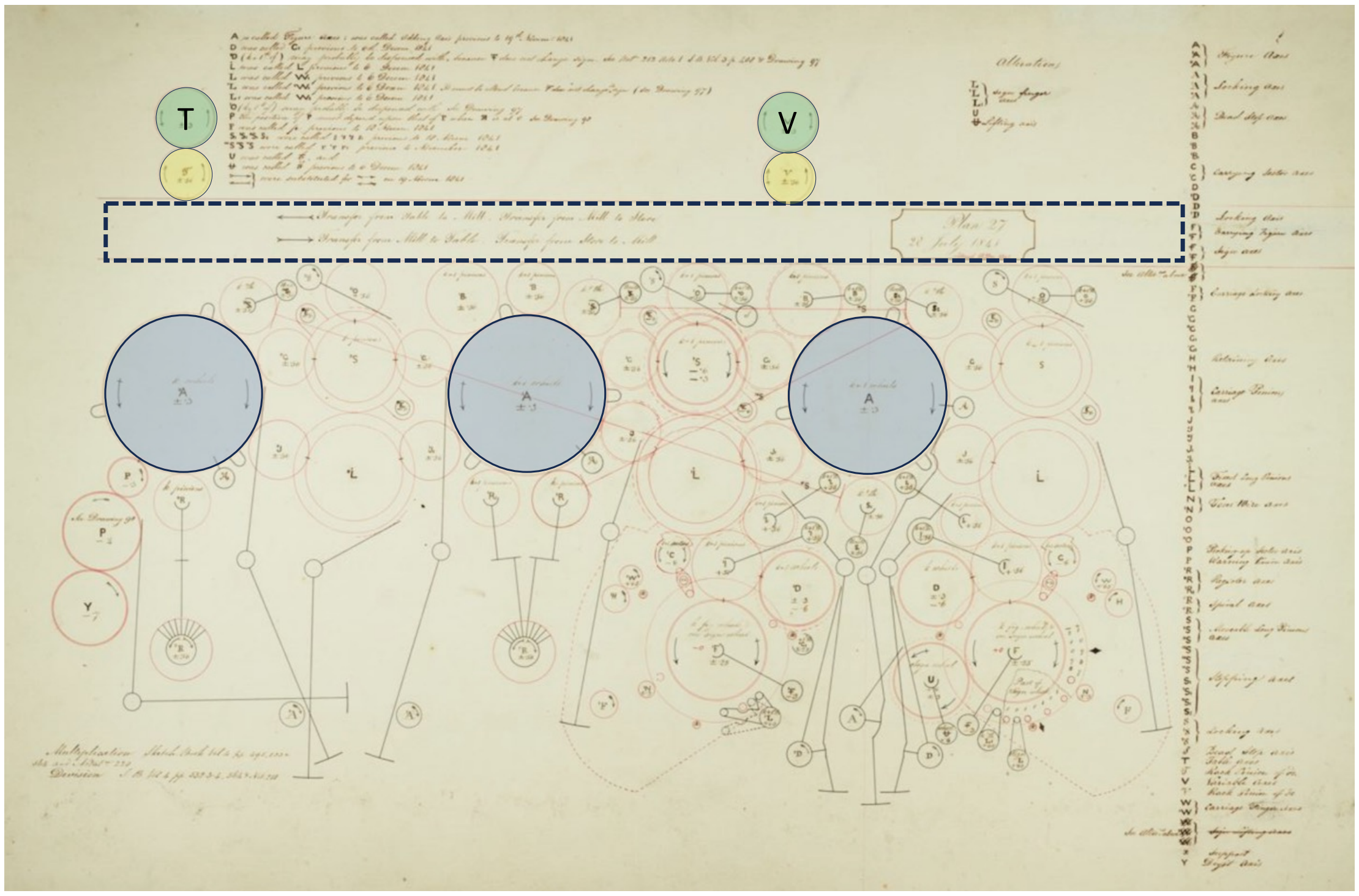

Fig. 11: Plan 26, AE only with linear buses (Babbage Papers)

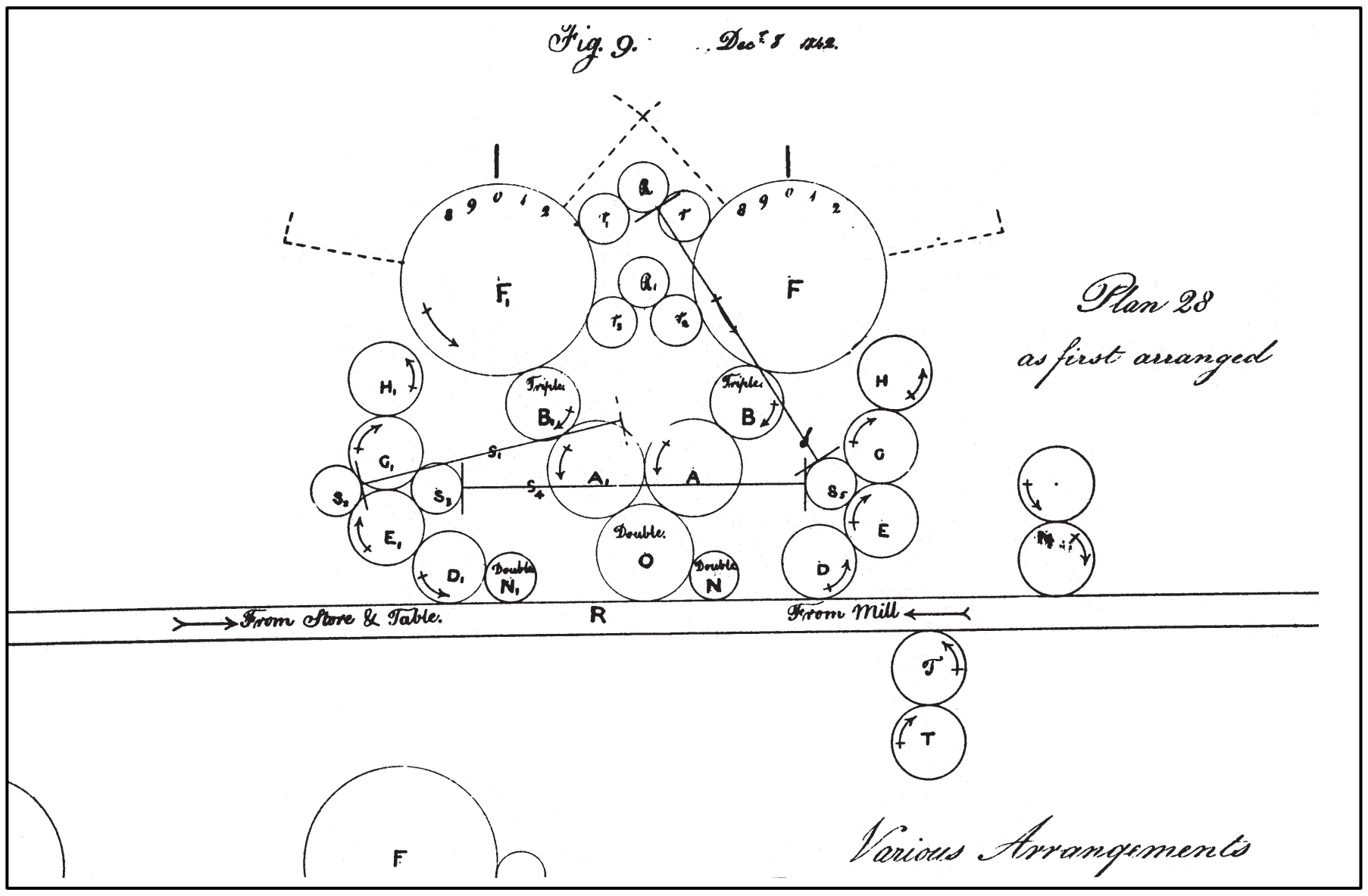

Fig. 12: Plan 28, based on a linear bus [Bromley 2000]

The last designs produced during the fifth phase of development of the AE were Plans 28 and 28a (Fig. 12). There is no simple comprehensive overview of those plans in the Babbage Papers. There are instead many views of subcomponents. In 2010, the “Plan 28 Project” was launched, with the objective of building a replica of the AE to mark its two hundredth anniversary in the 2030s. The project is still ongoing.[2]

## 6 The simplified demonstration model

Possibly realizing the monumental challenge of completing a "full size" AE, Babbage took to drafting at least two designs for significantly scaled-down models of the AE. The first, which he called a “General Plan of the Analytical Engine”, was sketched on October 2, 1866. Fig. 13 shows the new arrangement of parts, without specifying the control mechanisms.

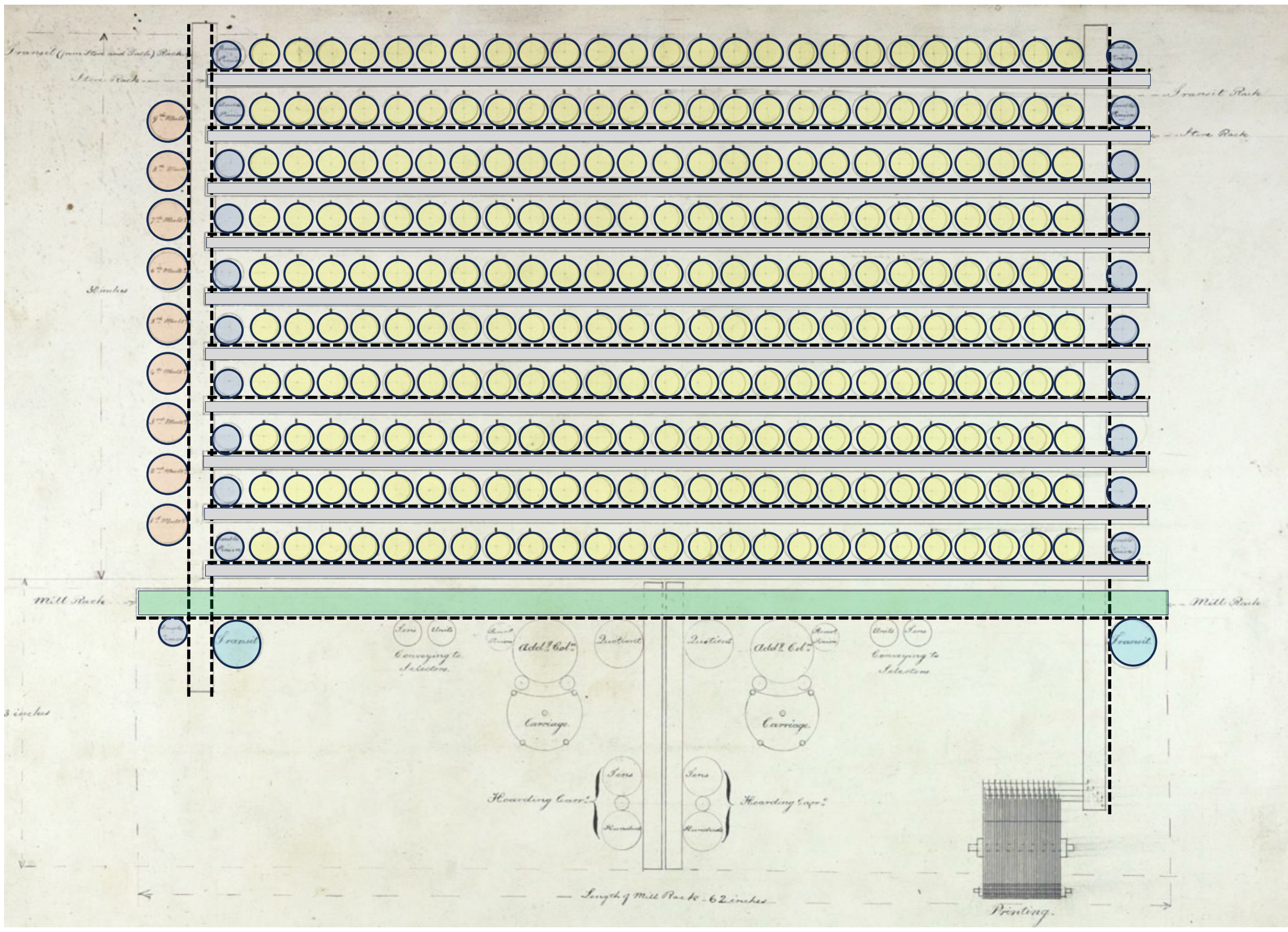

Fig. 13: General Plan of the Analytical Engine 1866, Babbage Papers, BAB/G/001

The most notable alteration lies in the realignment of the store along parallel racks (buses) that intersect vertical racks. The conversion of horizontal rack movement into vertical rack movement is facilitated by transit pinions (highlighted in blue). There are 10 horizontal buses for the store. The long horizontal bus highlighted in green is for the mill. The multiples of the multiplicand are stored in the nine axes shown on the left, in orange. They engage with the vertical rack to their right. As the reader can see, any displacement of a horizontal or vertical rack can be transmitted to any other rack, by means of the transit pinions.

[2] https://blog.plan28.org

The mill is perhaps the biggest surprise. It consists of two accumulators tailored for both addition and subtraction. Each of them has the associated machinery to implement anticipated carries. There are no intermediate gears to connect to the bus. Now, it is assumed that the axis that wants to transmit or send something (in the store, for example), lifts to engage with the racks, and lowers to disengage. While this approach eliminates a multitude of auxiliary gears, it places significant emphasis on ensuring precise angular positioning of the gears. Babbage, however, engineered an "interlocking mechanism" to facilitate such precise interfacing.

Another intriguing aspect is that this design appears to harken back to the mirrored mills of Plan 13. Each half of the mill seems to contain everything needed for addition and subtraction with an accumulator. The quotient can be stored in two axes and there are selectors on both sides of the mill for the numbers stored in the table of multiples (T1 to T9). There is also a provision for a hoarding carriage, one for each side. Hoarding carriages were used for the multiplication algorithm, because it took so much time, that anticipating carriages did not reduce the execution time significantly. It is possible that the two accumulators in the mill could be used together for multiplication and division. The absence of explicit control logic suggests that, in this demonstration model, the user himself would initiate operations and manually select the arguments, given the absence of a punched card reader. The size of the machine is 43 by 62 inches (109 by 157 cm). The model would have fit neatly on a table. Compare this with the dimensions estimated by Bromley for the AE of Plan 25: “The Analytical Engine would have stood about 15 feet high. The mill would have been about six feet in diameter and the store would have run 10-20 feet to one side” [Bromley 1998].

In the Babbage papers we find a second drawing of a “half size” AE of December 1866 (BAB/G/004), which includes only one of the two mills of the previous model and a handful of store variables. Notably, it also includes the mechanism for printing the results. One might envisage that Babbage intended to set up the contents of the store variables manually, and then show how the machine could compute the four arithmetic operations automatically, and even produce a printed result.

Two years prior to the creation of these models, Babbage had written that the culmination of the AE would become a venture destined for someone else in times yet to come: “Half a century may probably elapse before any one without those aids which I leave behind me, will attempt so unpromising a task. If, unwarned by my example, any man shall undertake and shall succeed in really constructing an engine embodying in itself the whole of the executive department of mathematical analysis upon different principles or by simpler mechanical means, I have no fear of leaving my reputation in his charge, for he alone will be fully able to appreciate the nature of my efforts and the value of their results” [Babbage 1864]. It is possible that Babbage’s son took the design for these models as the basis for his own prototype of the AE’S mill (Fig. 2). Bromley once lamented that nobody has taken that machine apart to examine exactly how it works [Grier 2004].

## 8 Programming the Analytical Engine

Babbage composed 27 code fragments for the AE that make the so-called software architecture of the machine understandable. The concept refers to the AE from the point of view of someone writing programs [Rojas 2017, 2021]. Fig. 14 shows the essential components that the programmer must keep in mind when writing code.

The sequence of control operations governing the mill is extracted from a bidirectional flow of punched cards. The two arguments for an arithmetic operation are fetched and put into A1 and A2. The result is written to R. The sequence of operand addresses is derived from two separate bidirectional strings of punched variable cards. The address of the first argument of an arithmetic operation is contained in the first string; the second argument is contained in the second string. Address 0 is for writing to A1 or A2, or for reading R from the mill. The carries (F, ‘F and ‘’F) of Plan 25 can be operated independently, for example during long multiplications, and work together as a kind of embedded difference engine.  Babbage also thought of number cards for reading numbers directly. These are not shown in this diagram.

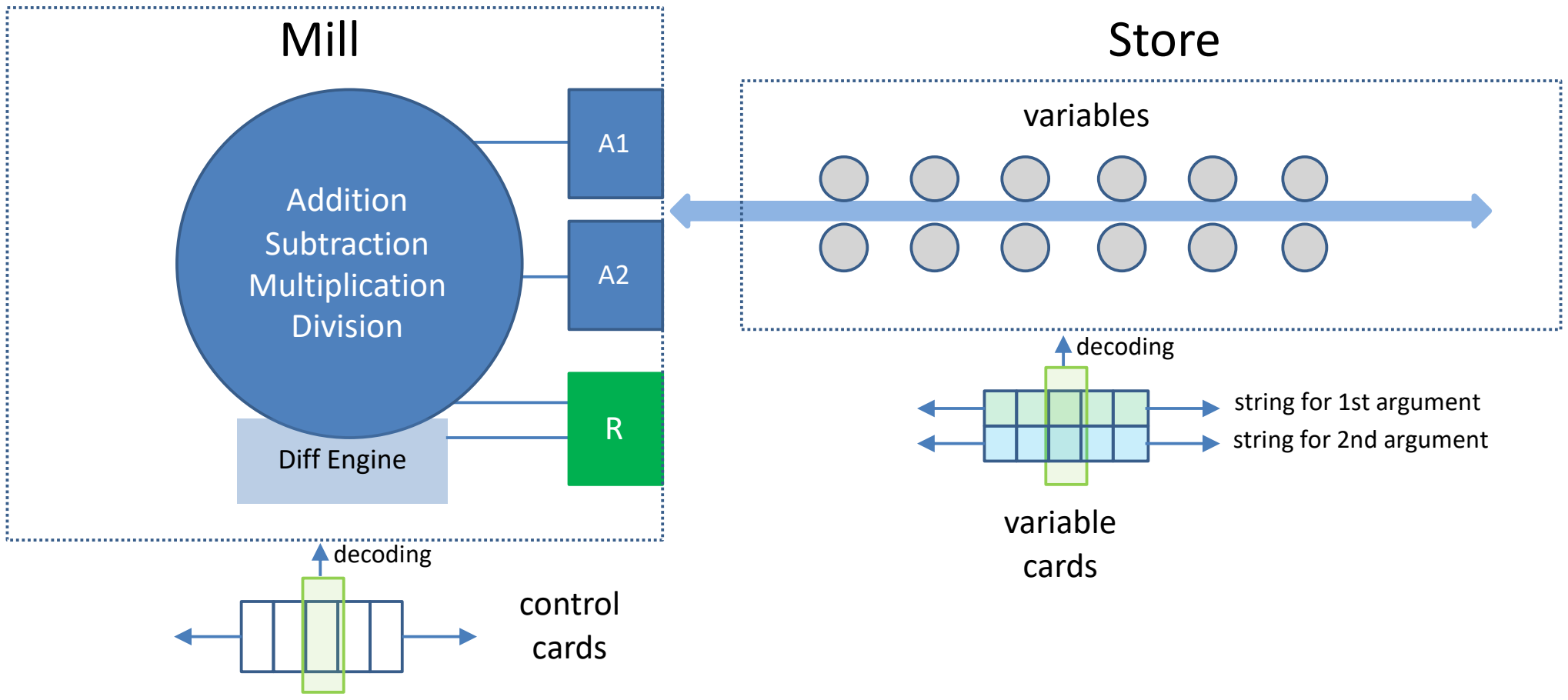

Fig. 14: The programming architecture of the Analytical Engine (up to 1840) with separate processor (mill) and memory (store).

According to the code written by Babbage, the mill was capable of executing the fundamental arithmetic operations. There was also a provision for carrying out conditional jumps within the stream of punched cards (using “combinatorial cards”), although it remains uncertain whether this included both backward and forward jumps [Bromley 2000]. Due to Babbage's separation of operation cards from argument cards, crafting static code (where operations are written alongside their arguments) is very difficult. To understand a program, one needs to execute it and generate a trace of the outcomes, in order to observe the dynamic interplay between operations and addresses coming from two separate streams. This stood as one of the primary limitations of Babbage's software architecture. Nevertheless, it is truly remarkable that we are able to discuss software that was written nearly two centuries ago.

In terms of the expected speed of the AE, Babbage wrote in [1864]: “Supposing the velocity of the moving parts of the Engine to be not greater than forty feet per minute, I have no doubt that: Sixty additions or subtractions may be completed and printed in one minute; One multiplication of two numbers, each of fifty figures, in one minute; One division of a number having 100 places of figures by another of 50 in one minute.” These optimistic projections stand in contrast to various other estimates that Babbage wrote on the margins of various drafts of the Analytical Engine.

## 9 Conclusions

Throughout this paper, we have traced the evolution of the Analytical Engine, as depicted in its different "Plans". The decision to separate the mill from the store was made very early in the design phase. Punched cards were incorporated into the design around 1836, and by 1840 Plan 25 had been drawn, a machine that incorporated the basic features of the AE, from the programmer’s point of view.

Subsequent phases of development introduced several key simplifications, including:

- The choice to do away with the shifting mechanism for the table axes T1 to T9, which stored multiples of the multiplicand. Instead of shifting these axes, the partial result of the multiplication (held in two accumulators) would be shifted.
- The adoption of non-restoring division, streamlining the division operation.
- The removal of intermediate gears in favor of lifting the axes to interface with other components and the racks.
- The elimination of the central wheel, in favor of multiple linear buses.

The simplest design was that adopted for the demonstration model of 1866 which would have fit on a table.

In terms of programming, Babbage wrote his last programs in 1840 and never came back to specify or revise the user instruction set. The only program that was likely authored by him and published after 1840 was the Bernoulli numbers program found in [Menabrea/Lovelace 1843], often credited to Ada Lovelace. Regarding this program Babbage stated: “We discussed together the various illustrations that might be introduced: I suggested several, but the selection was entirely her own. So also was the algebraic working out of the different problems, except, indeed, that relating to the numbers of Bernoulli, which I had offered to do to save Lady Lovelace the trouble. This she sent back to me for an amendment, having detected a grave mistake which I had made in the process” [Babbage 1864].

According to Bromley [Grier 2004] Babbage became so obsessed in refining the machine for speed and efficiency that he devoted excessive time to details rather than finalizing the overall design. When he finally returned to work on the AE, too much time had elapsed. However, the demonstration models could certainly have been built, and Babbage even considered the possibility of raising money through the demonstration of subcomponents. Yet, he recognized that he did not have much time left for such an endeavor: “On considering the whole question, I arrived at the conclusion, that to conduct the affair to a successful issue it would occupy so much of my own time to contrive and execute the machinery, and then to superintend the working out of the plan, that even if successful in point of pecuniary profit, it would be too late to avail myself of the money thus acquired to complete the Analytical Engine.” [Babbage 1864].

At the end, the AE was never finished. One can scarcely fathom the transformations that would have unfolded had the computer emerged in the 19th century rather than the middle of the 20th.